\documentclass[apj]{emulateapj}
\usepackage{natbib}
\usepackage{color}
\usepackage{graphicx}
\bibpunct{(}{)}{;}{a}{}{,}
\definecolor{atomictangerine}{rgb}{1.0, 0.6, 0.4}
\definecolor{beaver}{rgb}{0.62, 0.51, 0.44}
\definecolor{brightturquoise}{rgb}{0.03, 0.91, 0.87}

\begin{document}

   \title{
On the dearth of ultra-faint extremely metal poor galaxies 
}
\author{
J. S\'anchez Almeida\altaffilmark{1,2},
M.~E.~Filho\altaffilmark{1,2,3},
C. Dalla Vecchia\altaffilmark{1,2},
and 
E. D. Skillman\altaffilmark{4}
}

\altaffiltext{1}{Instituto de Astrof\'\i sica de Canarias, 38200 La Laguna, Tenerife, Spain}
\altaffiltext{2}{Departamento de Astrof\'\i sica, Universidad de La Laguna} 
\altaffiltext{3}{SIM/FEUP, 4200--465 Porto, Portugal}
\altaffiltext{4}{Minnesota Institute for Astrophysics, School of Physics and
Astronomy, University of Minnesota, Minneapolis, USA.}
\email{jos@iac.es}
\begin{abstract}
Local extremely metal-poor (XMP) galaxies are of particular astrophysical interest since they
allow us to look into physical processes characteristic of the early Universe, from the assembly 
of  galaxy disks to the formation of stars in conditions of low metallicity. Given the luminosity-metallicity 
relationship,  all galaxies fainter than $M_r\simeq -13$ are expected to be XMPs. 
Therefore, XMPs should be  common in galaxy surveys. However, they are not,
because several observational biases hamper their detection. This work compares the number of 
faint XMPs in the SDSS-DR7 spectroscopic survey with the expected number, given the known biases and 
the observed galaxy luminosity function.  The faint end of the luminosity function is poorly constrained 
observationally, but it determines the expected number of XMPs. 
Surprisingly, the  number of observed faint XMPs ($\sim 10$) is over-predicted by our calculation, 
unless  the upturn  in the faint end of the luminosity function is not present in the model. The lack of an 
upturn can be  naturally understood if most XMPs are central galaxies in their low-mass dark matter 
halos,  which are highly depleted in baryons due to interaction with the cosmic ultraviolet
background and to other physical processes. Our result also suggests that the upturn towards low 
luminosity of the  observed galaxy luminosity function is due to satellite galaxies.
\end{abstract}




\section{Motivation}\label{motivation}

Galaxies having a gas-phase metallicity smaller than a tenth of 
the solar metallicity are often known as extremely metal-poor
\citep[XMP; e.g.][]{2000A&ARv..10....1K}.  They are of astrophysical
interest for a number of reasons, among which include determining 
the primordial He abundance produced during  the 
Big Bang \citep{2010IAUS..268...91P,2016RvMP...88a5004C}, studying
star formation in conditions of low metallicity 
\citep{2014Natur.514..335S,2015Natur.525..218R,2015ApJ...805..145E,
2016ApJ...820..109F}, 
understanding the formation of dust in the early Universe \citep{2014Natur.505..186F}, 
analyzing primitive interstellar media \citep{2007ApJ...665.1115I}, 
constraining the properties of the first stars \citep{2005ApJS..161..240T,2015ApJ...801L..28K},
following the assembly of primitive disks 
\citep{2012ApJ...750...95E,2013ApJ...774...86E,2015ApJ...810L..15S,2016MNRAS.457.2605C}, 
and studying the intergalactic gas  \citep{2014A&ARv..22...71S,2014ApJ...783...45S}.

Unfortunately, the number of known XMPs remains small.
The review  paper by \citet{2000A&ARv..10....1K} contained only 31 XMPs,
\citet{2003ApJ...593L..73K} added 8 new targets from the early data 
release of the Sloan Digital Sky Survey (SDSS), the exploration in SDSS-DR6 by
\citet{2009A&A...505...63G} yielded 44 sources, 
and the systematic bibliographic search for all XMPs in 
literature and the SDSS-DR7 carried out by \citet{2011ApJ...743...77M} rendered 140 sources.
 \citeauthor{2011ApJ...743...77M} 
included targets found by \citet{2004ApJS..153..429K},
\citet{2004A&A...415...87I}, \citet{2006A&A...448..955I},
and \citet{2007ApJ...665.1115I}.
Although new local metal-poor objects have been discovered since 2011
\citep[e.g.,][]{2012A&A...546A.122I,2013AJ....146....3S,2015MNRAS.448.2687J,
2015A&A...579A..11G,2016ApJ...819..110S,2016ApJ...822..108H},   
XMPs remain uncommon.

The scarcity of known XMPs is in sharp contrast with the expectation that most 
galaxies are actually XMPs. The  problem has been put forward by several authors 
\citep[][]{2013AJ....146....3S,2013AJ....146..145M,
2015MNRAS.448.2687J,2016ApJ...819..110S}, 
with the following argument: there is a well-known relation between 
absolute luminosity or stellar mass and gas-phase metallicity 
\citep[e.g.,][]{1989ApJ...347..875S,2008ApJ...685..194S};
consequently all faint or low-mass galaxies should be XMPs. Using as 
the metallicity threshold,\footnote{If the solar oxygen abundance is 
$12+\log({\rm O/H})_\odot=8.69\pm0.05$
\citep{2009ARA&A..47..481A}, then the limit in 
Eq.~(\ref{faint4}) roughly 
corresponds to one tenth of the solar metallicity.} 
\begin{equation}
12+\log({\rm O/H})\leq 7.65,
\label{faint4}
\end{equation}
the metallicity versus absolute magnitude relationship 
by \citet{2012ApJ...754...98B} implies that galaxies  
with absolute $B$-band magnitude 
\begin{equation}
M_B \geq -12.5,
\label{faint}
\end{equation}
are XMPs. Similarly, the metallicity versus stellar mass relationship 
in the paper by \citet{2012ApJ...754...98B} 
entails that galaxies with stellar masses 
\begin{equation}
{\rm M}_\star \leq 1.1\times 10^7\,{\rm M}_\odot,
\label{faint2}
\end{equation}
correspond  to XMP galaxies. Equation~(\ref{faint}) can be re-written for the $r-$band magnitude
using the  transformation from $B$ to $r$  in \citet{2005AJ....130..873J} for typical colors of gas-rich 
galaxies  \citep[$g-r\simeq 0.4$; ][]{2009ARA&A..47..159B}, leading to 
\begin{equation}
M_r \geq -13.3.
\label{faint3}
\end{equation}
Since low-mass, low-luminosity galaxies outnumber, by far, high-mass, high-luminosity 
galaxies \citep[e.g.,][]{1988ARA&A..26..509B,
2009ARA&A..47..159B,
2014MNRAS.444.1647K},  most galaxies are expected to 
be XMPs. These faint XMPs, with magnitudes and/or masses below the thresholds
in Eqs.~(\ref{faint}), (\ref{faint2}), and (\ref{faint3}), will be designated
here as quiescent XMPs, or, QXMPs. The remaining XMPs are often denoted as
{\em active} XMPs. Note that, by definition, active XMPs are low-metallicity outliers of the luminosity-metallicity relationship. 

Hence, why are XMPs so unusual among the observed galaxies? 
It is argued that most QXMPs also have low surface brightness, so 
low as to be below the detection threshold of the largest surveys
\citep[see,][]{2013AJ....146....3S,2015MNRAS.448.2687J}. 

So far as we are aware, this qualitative argument has never been 
quantified. In other words, (1) what is the number of  QXMPs to be 
expected in current surveys? and, (2) is this prediction consistent with the 
number of known QXMPs? Our Paper addresses these two questions, 
finding that observational biases alone cannot account for the scarcity of
observed XMPs. The luminosity function of XMP galaxies must 
decline for objects fainter than the limits in  Eqs.~(\ref{faint}) and (\ref{faint2}).   
We show that such a drop is expected from cosmological numerical simulations, 
provided XMPs are central, rather than, satellite galaxies. The ultraviolet (UV) 
background  is then expected to prevent the formation of low-mass galaxies 
\citep[e.g.,][]
{1992MNRAS.256P..43E,1996ApJ...465..608T,2004ApJ...609..482K,2006Natur.441..322W,
2008MNRAS.390..920O}. 
Several physical processes may suppress gas accretion
and star formation in low-mass dark matter halos.
The cosmological UV background heats the intergalactic gas 
and establishes a minimum mass for halos that
can accrete gas. The gas in low-mass halos may also be 
photo-evaporated by the UV background after re-ionization. 
In addition, the ionizing radiation dissociates molecular hydrogen, 
which is the main coolant for low-metallicity gas, thus 
preventing star formation even before the gas is completely stripped
from the halos by other quenching processes. 
Stellar feedback processes, such as supernova explosions and stellar winds, are also 
able to remove gas from galaxy disks, reducing the star formation efficiency 
in low mass halos, where the gravitational binding energy is particularly low
\citep[e.g.,][]{1991ApJ...379...52W,2008MNRAS.387.1431D,2010Natur.463..203G,
2012RAA....12..917S,2014Natur.509..177V,2015MNRAS.446..521S}. 
Thus, when the galaxy has a stellar mass smaller than several $10^{8}\,M_\odot$, most of the gas 
that could be used to form stars returns unused to the inter-galactic medium  
\citep[e.g.,][]{2012MNRAS.421...98D,2012ApJ...760...50S,2014A&ARv..22...71S,2016ApJ...824...57C}.

The paper is organized as follows:
Section~\ref{obs_constraints} describes the number of QXMPs observed  
in the spectroscopic sample of the SDSS-DR7, which is chosen because this survey 
provides most of the known XMPs.
Section~\ref{estimate1} is devoted to estimating the expected number of QXMPs from 
the galaxy luminosity function, by firstly extrapolating the observed 
luminosity function to faint objects, and then 
including the effect of the baryon fraction changing with the dark matter
halo mass (Sect.~\ref{quenching}).
The results are discussed in Section~\ref{discussion}, including an analysis 
of the factors that limit the number of observed QXMPs, and how
they can be overcome in future searches (Sect.~\ref{exploring_limits}).
The number of observed QXMPs to be expected in other existing
and forthcoming surveys is determined in Section~\ref{other_surveys}.
The results are collected and summarized in Section~\ref{conclusions}. 
Throughout the paper, the Hubble constant $H_0$ is taken to be 70\,km\,s$^{-1}$\,Mpc$^{-1}$.

\section{Number of observed QXMPs in the SDSS-DR7}\label{obs_constraints}

The search for XMPs during the last decade has been very much
focused on the spectroscopic sample of the SDSS-DR7
\citep{2009ApJS..182..543A}. The purpose of this section is to evaluate how 
many  QXMPs have been found as part of these SDSS-DR7-based searches.
A comprehensive list is needed to compare the number of
observed and expected QXMPs. 

We have built the list by searching all recent papers that may have XMPs from 
the SDSS-DR7. The galaxies in this paper were filtered so as 
to keep only those from the SDSS-DR7 with metallicity and luminosity
below the thresholds in Eqs.~(\ref{faint4}) and (\ref{faint3}), respectively.
The samples that were analyzed are: 
\begin{enumerate}
\item The \citet[][ML+11]{2011ApJ...743...77M} XMP sample, 
which contains a compilation of all low metallicity ($12+\log[{\rm O/H}]\lesssim 7.65$) sources from 
the literature until the date of publication, plus several new targets from the SDSS-DR7 
spectroscopic sample. It uses photometry from the SDSS.
\item The \citet[][SA+16]{2016ApJ...819..110S}  
XMP sample from the SDSS-DR7,  with metallicity from SDSS spectra, and photometry also from the SDSS. 
\item The \citet[][Kara+13]{2013AJ....145..101K} 
sample,  which is the latest version of the nearby galaxy reference catalog.  
Only sources with $M_B \geq -12.5$~mag are considered 
here,  where both the metallicity and photometry are compiled from literature.
\item The \citet[][Berg+12]{2012ApJ...754...98B} 
sample, a subsample of low-luminosity galaxies within 11~Mpc, with metallicities and photometry from
 Multiple Mirror Telescope (MMT) observations.
\item The \citet[][Izo+12]{2012A&A...546A.122I} 
sample, a study of metal-poor emission-line SDSS-selected galaxies, with metallicities from Apache Point Observatory 
(APO) 3.5 m and/or MMT, and photometry from the SDSS.  
\item The \citet[][James+15]{2015MNRAS.448.2687J} 
sample, a set of SDSS-selected blue diffuse dwarf galaxies, with metallicities from MMT
observations and photometry from the SDSS.
\end{enumerate}

From these six samples, we selected galaxies fulfilling the following criteria:  
(a) appear in the SDSS-DR7 spectroscopic catalog, 
(b) their $B$-band absolute magnitudes are larger than -12.5, and 
(c) have metallicities $12+\log({\rm O/H}) \leq 7.65$. 
For the resulting sources, 
we checked whether the photometry was reliable, particularly in those cases where only 
SDSS photometry is available. For the SDSS photometry, we first visually inspected the SDSS images 
of the sources and registered the source size. The visual sizes were then compared with the SDSS 
Petrosian g-band radius at 90\% of the light, obtained at the position of the SDSS spectroscopic target. 
If the sizes were well-matched, the SDSS photometry was deemed reliable, and the source was 
retained as a QXMP. If the sizes were not well-matched, we then looked into literature for a value 
for the photometry. 
Most of them happen to have alternative photometry, which allowed us to exclude 90\,\%~ of them 
as QXMPs. Only three objects had no alternative photometry, and they were discarded 
by the argument that their chance of being an QXMP is the same as those with photometry. 
This chance  is only 10\,\%, which amounts to 0.3 sources.  

\begin{figure*}
\includegraphics[width=0.48\textwidth]{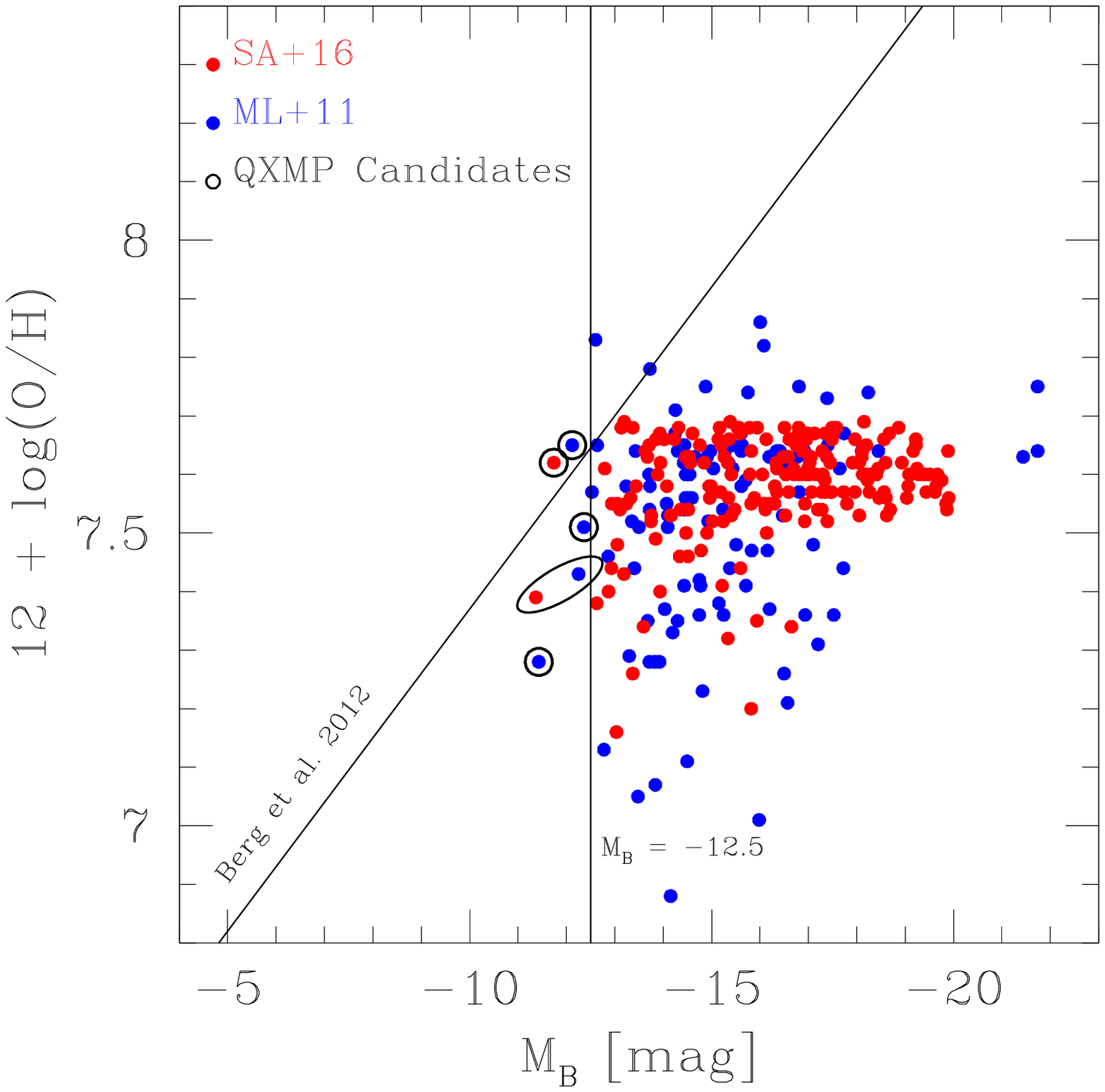}
\includegraphics[width=0.48\textwidth]{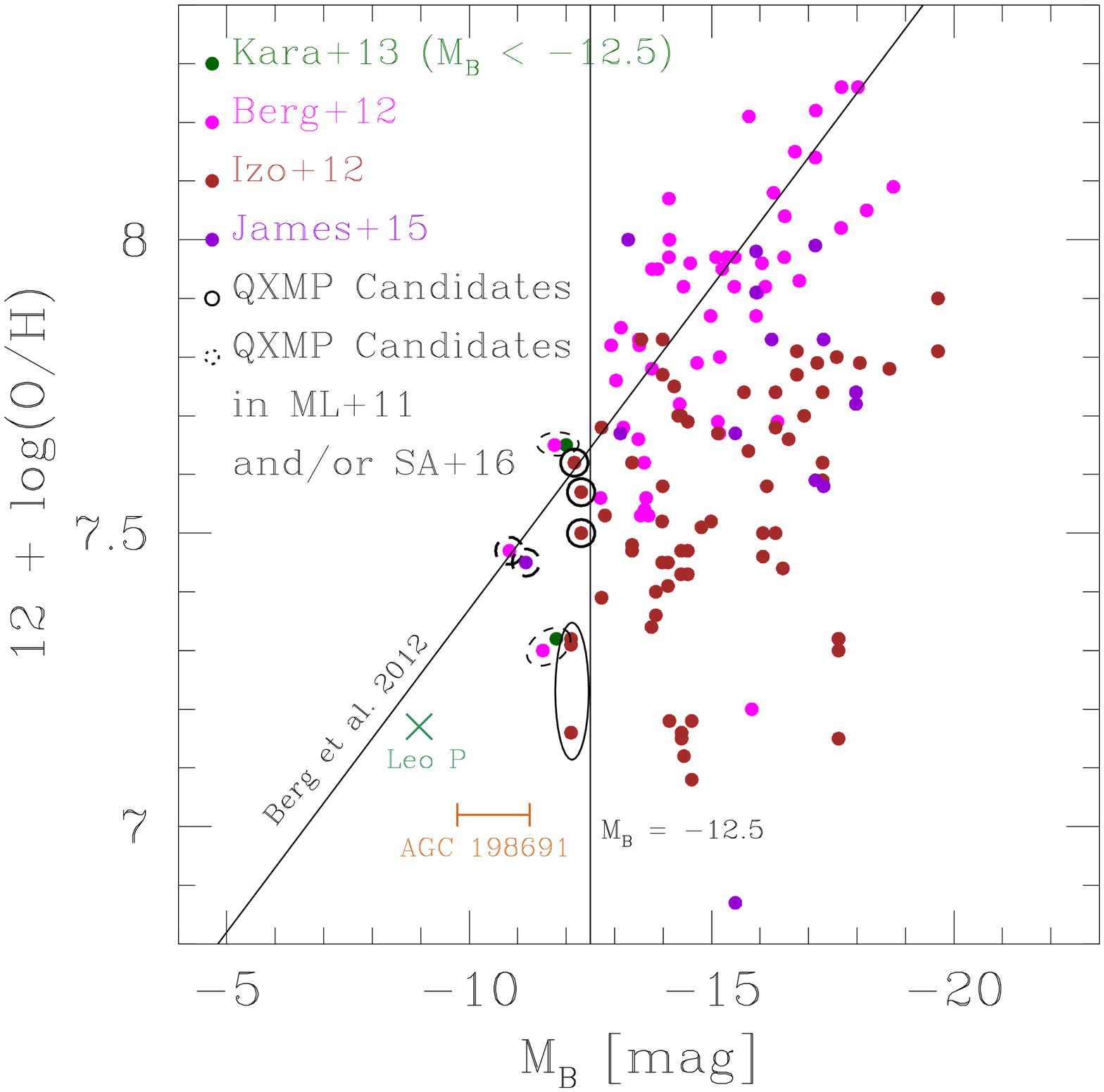}
\caption{Oxygen abundance vs absolute $B$ magnitude for galaxies 
found in recent searches containing XMPs. We are interested 
in those objects that are QXMPs (i.e., to the left of the vertical black solid 
lines, $M_B = -12.5$) and which simultaneously appear in the SDSS-DR7 
spectroscopic sample.  The luminosity-metallicity  relationship by \citet{2012ApJ...754...98B}, 
used to define QXMPs,  is included.
(a) Compilation by \citet[][blue symbols]{2011ApJ...743...77M} of all XMPs in the literature
up to the date of publication, and 
high-electron temperature XMPs from the SDSS-DR7  by 
\citet[][red symbols]{2016ApJ...819..110S}.  Only the five encircled 
objects are  bona-fide QXMPs in the  SDSS-DR7 spectroscopic sample.
One of these appears in the two samples with slightly different magnitude and metallicity; 
the two corresponding points are encircled together.
(b) Objects from \citet[][green symbols]{2013AJ....145..101K},  
\citet[][brown symbols]{2012A&A...546A.122I},
\citet[][purple symbols]{2015MNRAS.448.2687J}, and 
\citet[][magenta symbols]{2012ApJ...754...98B}. 
The prototype QXMP Leo~P  \citep{2013AJ....146....3S} 
and the recently discovered AGC~198691
\citep{2016ApJ...822..108H} are also included in the figure for reference,
even though they are not part of the SDSS-DR7 spectroscopic sample. 
Four new QXMPs have been identified, and are marked with a solid outline. One of
the four  (J1056+36) has   
three HII regions having slightly different abundances.
The objects with the dashed outlines 
are QXMPs already included in (a) --two of these have two
different abundance estimates so that the dashed 
outline is elongated.
}
\label{observation}
\end{figure*}

Figures  \ref{observation}a and \ref{observation}b summarize the selection procedure
and its outcome. They show oxygen abundance versus absolute $B$ 
magnitude for the galaxies in the previous references. In order to avoid overcrowding,
the samples are split into  two plots, and only XMPs with reliable magnitudes are shown 
for $M_B > -12.5$. 
Figure~\ref{observation}a contains the  compilation ML+11, plus 
the XMPs from the SDSS-DR7 recently identified in SA+16. 
There are only five targets with $M_B > -12.5$ that seem to be bona-fide QXMPs in the 
SDSS-DR7 spectroscopic database. Those are encircled with black outlines in Fig.~\ref{observation}a.
One of these appears in the two samples with slightly different magnitude and metallicity; 
the two corresponding points are encircled together.

Figure~\ref{observation}b contains  the low-luminosity objects with measured metallicity 
from the four remaining samples: Kara+13 (green symbols), Izo+12 (brown symbols),
James+15 (purple symbols), and Berg+12 (magenta symbols).
The galaxies in Berg+12 are those used to derive the luminosity-metallicity relationship 
leading to the  QXMP magnitude limit in Eq.~(\ref{faint}); the relationship is shown as a slanted 
black line in Figs.~\ref{observation}a and \ref{observation}b. 
The prototype QXMP Leo~P  \citep{2013AJ....146....3S} and the recently discovered 
AGC~198691\footnote{The distance to AGC198691 is unknown. We adopted a range between 8 and 16 Mpc, 
as in \citet{2016ApJ...822..108H}, to estimate the range of absolute magnitudes shown in 
Figs.~\ref{observation} and \ref{sbvsmag}. 
The surface brightness has been obtained 
assuming an angular diameter of 8.1~arcsec.} 
\citep{2016ApJ...822..108H} are also included in the figure for reference,
even though they do not belong to  the SDSS-DR7 spectroscopic sample. 
After a screening similar to the one carried out with the objects in 
Fig.~\ref{observation}a, we are left with only four new targets outlined
with solid lines in  Fig.~\ref{observation}b. 
The samples represented in Fig.~\ref{observation}b contain 4 additional QXMPs, 
which are outlined with dashed lines in the figure.  They are already in the samples
by ML+11 or SA+16, and so, they do not contribute to the total number of QXMPs.

\begin{figure*}
\includegraphics[clip,width=18cm]{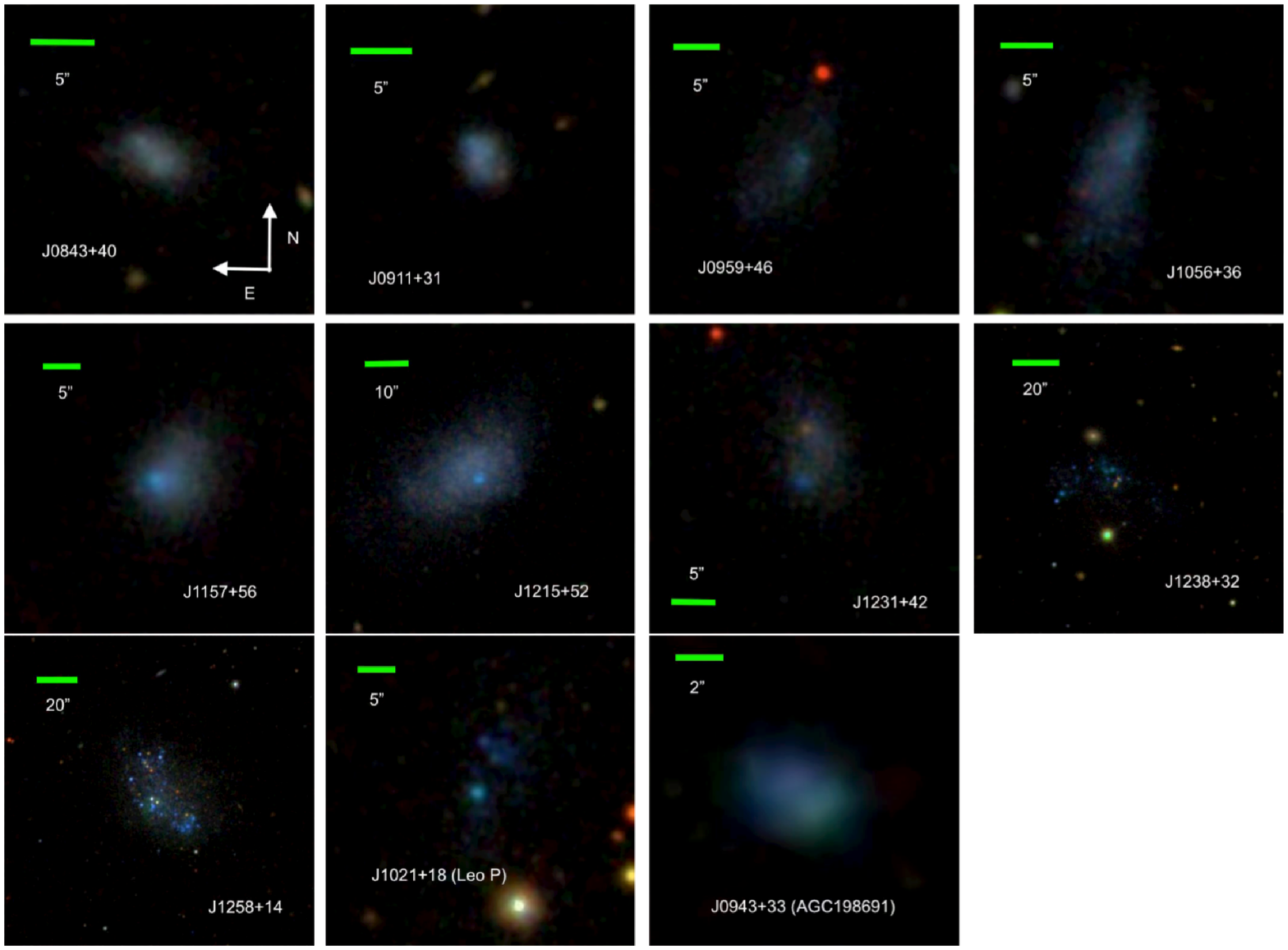}
%
\caption{
Images of the nine QXMPs in the SDSS-DR7 spectroscopic survey, 
with the names given in the insets. Their properties are listed in 
Table~\ref{coordinates}.
Although the prototype QXMP Leo~P is not in the SDSS-DR7 spectroscopic survey,
it is included for reference. The same happens with AGC~198691. 
The orientation  of all images is given by the 
arrows in the first image, with the angular size shown in the individual images by the green scale.
}
\label{images}
\end{figure*}

\begin{figure}
\includegraphics[width=0.45\textwidth]{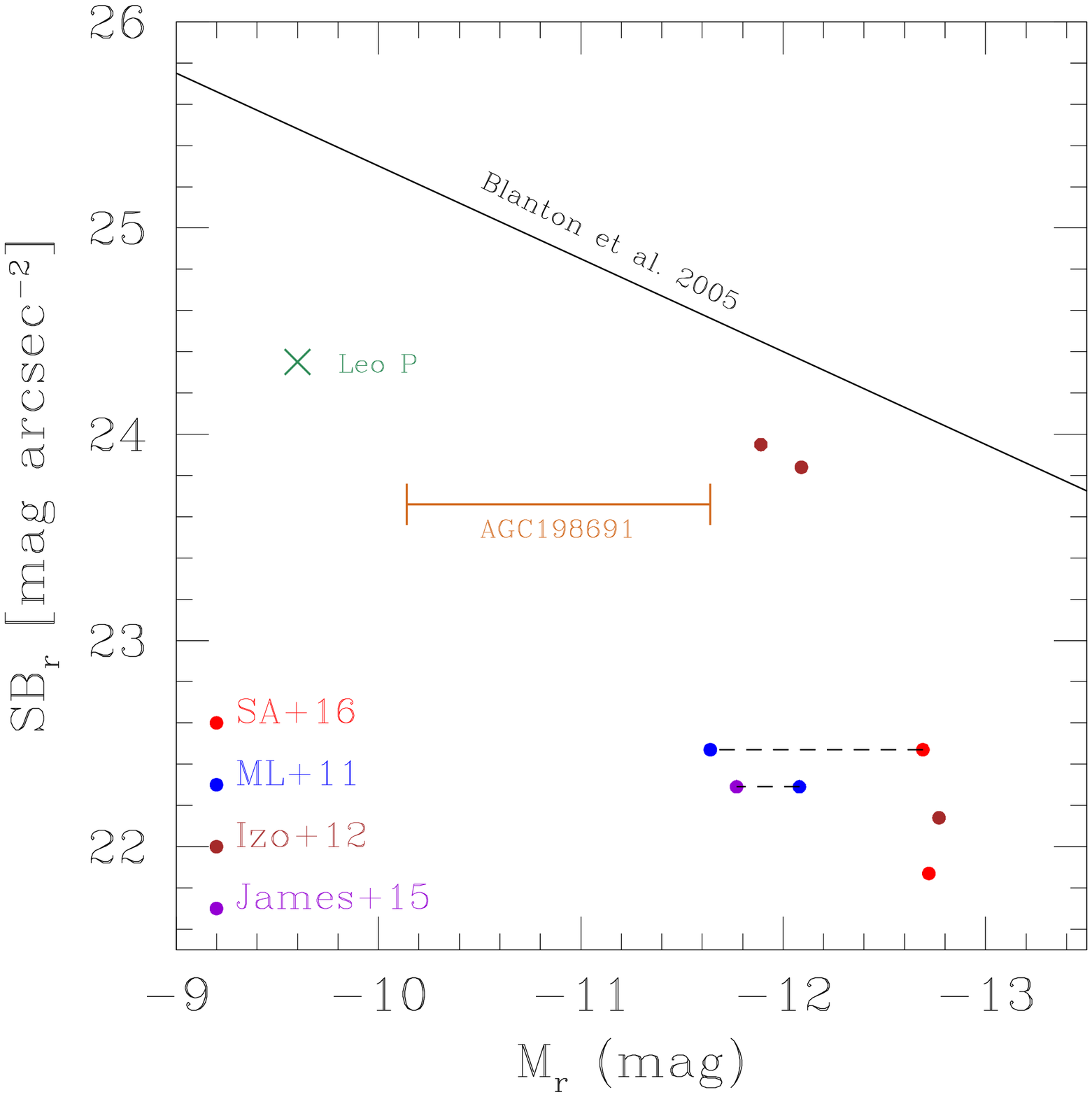} 
\caption{$SB_r$ versus $M_r$ for the QXMP with spectra in the SDSS-DR7 and reliable SDSS photometry.
Leo P and AGC~198691 are also included for reference. 
Note that they are high-SB outliers of the $SB_r$ versus $M_r$ relationship found 
for brighter objects, 
and extrapolated to fainter galaxies in this plot (the black solid line).
The pairs of points joined by dashed lines correspond to the same object 
reported in two different samples. 
}
\label{sbvsmag}
\end{figure}

At the end of our selection process, we are left with nine QXMPs, i.e.,
\begin{equation}
N^{\rm obs}_{\rm QXMP}=9\pm 3,
\label{obs_qxmp}
\end{equation}  
where the error considers only the Poissonian statistical error  
associated with the process of counting \citep[e.g.,][]{martin71}. 
It is important to realize that Eq.~(\ref{obs_qxmp}) probably represents 
an upper limit. Since QXMPs are so uncommon, the chances of having false 
positives (non-QXMP galaxies misidentified as QXMP) are 
expected to be much greater than the chances of having false negatives 
(QXMPs excluded by mistake). The coordinates and main properties of the nine QXMPs
%
are listed in Table~\ref{coordinates}, 
while their SDSS images are shown in Fig.~\ref{images}. 
Many of them are contained transversally in more than 
one of the analyzed samples. 
\begin{table*}
\begin{centering}
\caption{Bona-fide QXMPs in the SDSS-DR7 spectroscopic survey.}
\begin{tabular}{ccccccccc}
\hline
Name\,$^a$           & $12+\log({\rm O/H})$ & $M_B$ & $M_r$  & $SB_r$ &D\,$^b$& Reference & Also in  \\
&&&& [mag arcsec$^{-2}$]& [Mpc] &\,$^c$&\,$^c$\\
\hline
J084338.0+402547.1 & $7.57\pm 0.06$  & -12.3 & -12.8 & 22.1 & 10.2 &Izo+12 & -- \\                                
J091159.4+313534.4&$7.51\pm 0.14$&-12.4&-12.7&21.9&11.6&ML+11& -- \\
J095905.7+462650.5 & $7.50\pm 0.05$  & -12.3 & -11.9 & 24.0 & 8.0& Izo+12 & -- \\  
J105640.3+360827.9&$7.26\pm 0.09$&-12.1&--&--&9.2&Kara+13 ($M_B$)& Izo+12 ($12+\log[{\rm O/H}])$\\                              
J115754.2+563816.7 & $7.62\pm 0.11$  & -11.7 & -12.1 & 22.3 & 5.8 & SA+16 & Kara+13, James+15 \\
J121546.6+522313.8 & $7.39\pm 0.15$  & -11.4 & -11.6 & 22.5 & 2.1 & SA+16 & ML+11,  Berg+12\\
J123109.1+420533.9 & $7.62\pm 0.04$  & -12.2 & -12.1 & 23.8 & 8.2 &Izo+12 & Kara+13  \\ 
J123839.1+324555.9&$7.28\pm 0.07$&-11.4&--&--&3.1&ML+11&Kara+13, Berg+12\\
J125840.1+141308.1&$7.65\pm 0.06$&-12.1&--&--&2.2&ML+11&Kara+13, Berg+12\\
\hline
\end{tabular}
\raggedright
\begin{tabular}{l}
$^a$\,The name includes the coordinates RA and DEC.\\
$^b$\,Distance adopted in the respective reference to determine absolute magnitudes.\\
$^c$\,\citet[][ML+11]{2011ApJ...743...77M}, {\citet[][Izo+12]{2012A&A...546A.122I}}, 
\citet[][Berg+12]{2012ApJ...754...98B}, \\
~~\citet[][Kara+13]{2013AJ....145..101K}, \citet[][James+15]{2015MNRAS.448.2687J},
\citet[][SA+16]{2016ApJ...819..110S}.\\
\end{tabular}
\label{coordinates}
\end{centering}
\end{table*}
%
%

As we argue in Sect.~\ref{motivation}, QXMPs are expected to have  low surface 
brightness (SB). Figure~\ref{sbvsmag} shows the SB versus absolute magnitude in the $r$-band
for the QXMPs with reliable SDSS photometry. 
All surface brightnesses referred to in the paper are half-light surface brightnesses.
Despite their low SB (22 -- 24~mag~arcsec$^{-1}$; see Fig.~\ref{sbvsmag} and Table~\ref{coordinates}),
they tend to be  brighter than expected from extrapolating the relationship 
between  magnitude and SB found for brighter 
galaxies \citep[the black solid line in Fig.~\ref{sbvsmag}, from ][]{2005ApJ...631..208B}. 
Only two of the targets appear to follow the relationship (J0959+46 and J1231+42; 
compare their images with Leo~P in Fig.~\ref{images}).  
The origin of this unexpectedly large SB  can be pinned down to the bias of 
the observation toward high-SB objects, and the intrinsic scatter in the 
relationship between SB and magnitude. Given an absolute magnitude,
observations preferentially pick up the objects of largest SB.  This scatter
is important for estimating the expected number of QXMPs, and so it is 
discussed and treated in Sect.~\ref{error} and the Appendix.

\section{Number of faint XMPs to be expected in the SDSS-DR7}\label{estimate1}

\subsection{Calculation of the expected number of QXMPs}

\begin{figure}
\includegraphics[width=0.49\textwidth]{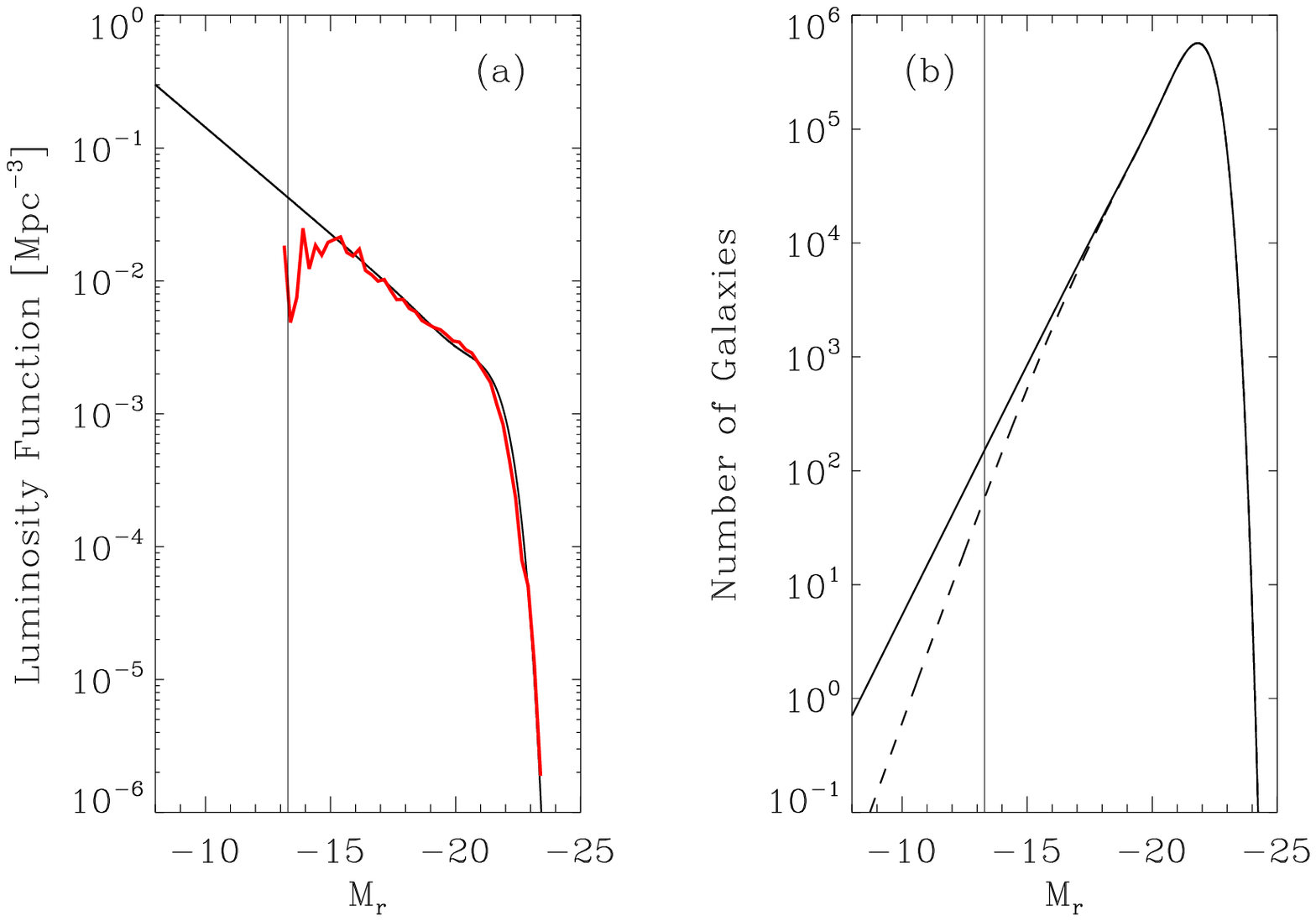}
\caption{(a) Galaxy LF in the $r$-band from 
\citet{2005ApJ...631..208B}. The red line shows the actual data, whereas the black solid line
represents a double Schechter function fit, which we use to describe the number 
of QXMPs with  magnitudes to the left of the vertical black solid line.
(b) Number of galaxies per unit magnitude to be expected from the 
SDSS-DR7 spectroscopic  survey.  
The black solid line shows the effect of the Malmquist bias 
(apparent magnitude limit cutoff), whereas the black dashed line also 
includes the incompleteness for low-SB objects.
}
\label{fig1}
\end{figure}

The galaxy luminosity function (LF), $\Phi(M)$, is defined as the number of galaxies with absolute magnitude $M$,
per unit volume and unit magnitude. Figure~\ref{fig1}a shows the LF in the $r$-band  determined by 
\citet{2005ApJ...631..208B} from the SDSS-DR2 data.  The number of galaxies in a survey  fainter 
than a given limiting magnitude, $M_{lim}$, is 
\begin{equation}
N(M > M_{lim})=\int^{M_{1}}_{M_{lim}} S(M)\,\Phi(M)\,dM,
\label{eq1}
\end{equation}
with $S(M)$ the selection function that provides the effective 
volume that is sampled by the survey. Equation~(\ref{eq1}) assumes $S$ to 
depend only on the absolute magnitude of the galaxy, which is reasonable
in our case, and simplifies the treatment. This assumption will be 
relaxed  later on. Equation~(\ref{eq1}) also assumes
the existence of a limit for the magnitude of the faintest galaxy, $M_{1}$.

In the case of a volume-limited sample, $S$ is just the volume 
of the sample, and so independent of the absolute magnitude. In  the
case of an apparent magnitude-limited survey, where all galaxies brighter
than the apparent magnitude, $m_{lim}$, are included,
$S(M)$ is the volume where galaxies of magnitude $M$ have
apparent magnitude $m_{lim}$ or brighter, i.e., 
\begin{equation}
S(M)=V(M)={{d^3}\over{3}}\Omega,
\label{pi_eq7}
\end{equation}
with $\Omega$ the solid angle covered by the survey, and 
$d$ the maximum distance at which a galaxy of magnitude $M$
can be observed, i.e.,
\begin{equation}
\log(d)= {{1}\over{5}}(m_{lim}-M)-5,
\end{equation}
with $d$ in Mpc.
The SDSS spectroscopic survey was designed to be apparent magnitude-limited,
with a limit in the $r$-band given by
\begin{equation}
m_{lim}=17.77.
\label{m_limit}
\end{equation}

\begin{figure}
\includegraphics[width=0.48\textwidth]{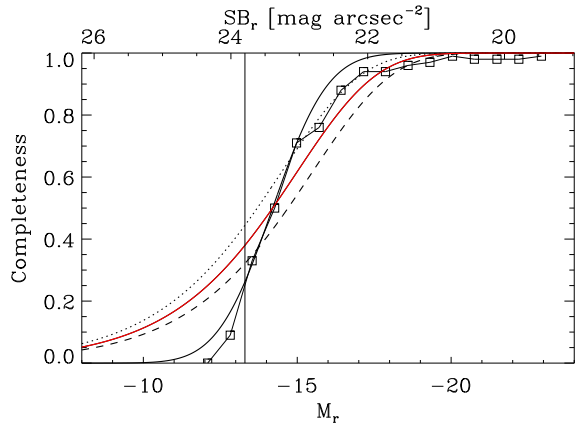}
\caption{Completeness function for the SDSS due to the low SB
of the targets. The symbols joined by a solid line combine the 
total completeness estimated by  \citet{2005ApJ...631..208B}
with the relation between $SB_r$ and $M_r$ in  Eq.~(\ref{mag_mag}). 
This completeness function is  approximated in the analysis by the erf
function 
shown as a  thick solid  line. The red solid line is the effective completeness function
resulting from the scatter in the $SB_r$ vs $M_r$ relationship.
The black dotted and dashed lines are the red solid line shifted by plus and minus 
half a magnitude, respectively, and encompass the statistical error of the 
completeness function. 
The vertical black line, $M_r=-13.3$, marks the QXMP limit.
The axis on top gives the $SB_r$ in mag~arcsec$^{-2}$, corresponding to 
the absolute magnitudes in the bottom axis.}
\label{completeness}
\end{figure}

In practice, all surveys have a limit in surface brightness. 
\citet{2005ApJ...631..208B} work it out for the SDSS, showing 
that the completeness of the spectroscopic survey decreases 
drastically for an average 
surface brightness, $SB_{r}$, fainter than 23~mag~arcsec$^{-2}$,
reaching only 10\,\%\ completeness at  24~mag~arcsec$^{-2}$. 
This bias against low-SB objects is particularly severe for QXMPs. 
Galaxies fainter than the limit in Eq.~(\ref{faint}) may, in principle, 
have any surface brightness. However, faint galaxies tend to have a low SB
as well \citep[e.g.,][]{1985ApJ...295...73K,1999ASPC..170..169S}. 
\citet{2005ApJ...631..208B} give the following relationship between
surface brightness and absolute magnitude in the $r$-band, 
\begin{equation}
SB_r=23.8+0.45\,(M_r+13.3) .
\label{mag_mag}
\end{equation} 
This relationship is in close agreement with others found in literature
\citep[e.g.,][]{1985ApJ...295...73K,2012AJ....143..102G}, 
and appears to be valid down to very low magnitudes 
\citep[even for $M_B > -8$; e.g.,][]{2013AJ....145..101K}.
Equation~(\ref{mag_mag}) implies that galaxies fainter than the QXMP limit
(Eq.~[\ref{faint3}]) are fainter than 23.8 mag~arcsec$^{-2}$, and so, 
potentially subject to
severe incompleteness in the SDSS. Completeness is quantified using 
the completeness function, which gives the fraction of galaxies with a given SB 
that are detected in the survey, $C'(SB)$. 
Combining the SDSS completeness function 
by \citet[][Fig.~3]{2005ApJ...631..208B} with Eq.~(\ref{mag_mag}), one 
finds  the completeness function, $C(M)$, in Fig.~\ref{completeness}
(the symbols joined by a solid line).  Including completeness,  
the selection function turns out to be 
\begin{equation}
S(M)=V(M)\,C(M),
\label{select1}
\end{equation}
which remains a function of the absolute magnitude only.
As we show in the Appendix, Eq.~(\ref{select1}) remains formally valid 
when the scatter of the relationship between $SB$ and $M$ is taken into 
account.  This scatter is bound to be important in the analysis, since most
observed QXMPs are high-SB outliers of the $SB_r$ versus $M_r$ relationship
(Fig.~\ref{sbvsmag}). 
In this case, the completeness, $C(M)$, in Eq.~(\ref{select1}) must be replaced with an
effective completeness, 
\begin{equation}
S(M)=V(M)\,C_{eff}(M),
\label{eq:select}
\end{equation}
where
\begin{equation}
C_{eff}(M)=\int_{\forall\,SB} P(SB|M)\,C'(SB)\,dSB.
\label{eq:ceff}
\end{equation} 
$P(SB|M)$ stands for the conditional probability function of having a surface brightness
$SB$ when the magnitude of the galaxy is $M$. $C'(SB)$ represents the completeness
function in terms of $SB$. $P(SB|M)$ is also provided by \citet{2005ApJ...631..208B} (see
Appendix \ref{appa}), and the resulting $C_{eff}(M)$ is represented in Fig.~\ref{completeness} as the red solid line.
In order to evaluate the integral in Eq.~(\ref{eq:ceff}), we use an erf function fit 
to the actual discrete 
completeness measured by  \citet{2005ApJ...631..208B} (i.e., we use the smooth thick solid line in 
Fig.~\ref{completeness},  meant to reproduce the symbols).
As can appreciated in Fig.~\ref{completeness}, the scatter increases the effective completeness 
of the  survey at low SB, and this occurs  because 
many targets happen to have a SB larger than that assigned by Eq.~(\ref{mag_mag}), and those 
are the ones that are preferentially selected.

The integrand of Eq.~(\ref{eq1}) gives the number of galaxies of a given magnitude to 
be expected in a survey.   Figure~\ref{fig1}b shows this integrand for the SDSS-DR7 
spectroscopic survey, which implies employing the $m_{limit}$ in Eq.~(\ref{m_limit}), 
and $\Omega= 2.45\,$srad \citep{2009ApJS..182..543A}. We use  the LF for extremely
low-luminosity galaxies determined by  \citet{2005ApJ...631..208B}. The actual
LF is shown as the red solid line in Fig.~\ref{fig1}a, although 
we use for the calculations  the double Schechter function fitted to the observations
by \citet{2005ApJ...631..208B}, 
i.e., 
\begin{equation}
\Phi(M)=(0.4\,\ln 10\,h^3\,{\rm Mpc}^{-3})\,{\Large\exp}\big[-10^{-0.4\,(M-M_\star)}\big]\,\times 
\end{equation}
\begin{displaymath}
\big[\phi_{\star,1}\,10^{-0.4\,(M-M_\star)*(1+\alpha_1)}+\phi_{\star,2}\,10^{-0.4\,(M-M_\star)*(1+\alpha_2)}\big]
\end{displaymath}
with $\phi_{\star,1}=0.0134$, $\phi_{\star,2}=0.0086$, $\alpha_1=0.33$, $\alpha_2=-1.40$, 
$M_\star=-19.99-5\,\log h$,
and $h$ is the Hubble constant normalized to 100 km\,s$^{-1}$\,Mpc$^{-1}$.
This   double Schechter function is shown as the black solid line in Fig.~\ref{fig1}a. 
Thus, our estimates are based on the extrapolation to low luminosity of the 
observed LF, which yields a function that
continues growing towards the region occupied by the QXMPs (to the 
left of the vertical black line in Fig.~\ref{fig1}a). Assuming the Malmquist bias alone,
i.e., using Eq.~(\ref{pi_eq7}) for the selection function, the number of 
expected QXMPs, $N_{\rm QXMP}$, is 149 (see Table~\ref{table1}). This
figure results from integrating the solid line in Fig.~\ref{fig1}b up to
the QXMP threshold. 
When the full selection function is considered  (Eq.~[\ref{eq:select}]),
the total number of expected QXMP galaxies turns out to be
\begin{equation}
N_{\rm QXMP}\simeq 42.
\label{nqxmp}
\end{equation}
We integrate from $M_r=-13.3$ to $-8$. 
The upper limit is taken from \citet[][Fig.~10]{2013AJ....145..101K} as the 
faintest magnitude of the late-type  galaxies in the local Universe. 
This upper limit, however,  does  not affect $N_{\rm QXMP}$ since the selection 
function is very small at low luminosities.  
Even though the number in Eq.~(\ref{nqxmp}) represents a minuscule 
fraction of the galaxies to be expected in the full survey 
(1.1$\times 10^6$; Table~\ref{table1}), the actual number exceeds the 
$9\pm 3$  QXMP galaxies that are observed (Sect.~\ref{obs_constraints}). 
The predicted $N_{\rm QXMP}$ depends on the completeness function, and
it falls off towards low SB, which is uncertain. We work out the error budget  in Sect.~\ref{error},
yielding a number of expected QXMPs between 12 and 73, with the high value
range strongly favored.  
Thus, the discrepancy between observations and predictions remains 
even when  uncertainties  are taken into account. 

Our detailed description of differences between the number of observed and 
predicted  QXMPs should not override the fact that these two numbers are always 
very small. QXMPs are, at most, a few tens in a survey such as the SDSS-DR7, 
which contains almost one million galaxies with spectra. QXMPs outnumber any 
other type of galaxy, but only a tiny fraction of  these is detected:
in the prediction described above, 87\,\% of the galaxies are QXMPs, but they 
constitute only 0.004\,\% of the detected galaxies  (Table~\ref{table1}). 
The actual percentages depend on the specific assumptions   (see Table~\ref{table1}), 
but the vast disproportion between the true number of QXMPs and their paucity in 
surveys always holds true.

\begin{table*}
\begin{centering}
\caption{Number of galaxies to be expected in a SDSS-DR7-like survey.}
\begin{tabular}{ l  c  c }
\hline
Description & Number or Percentage & Comment\\
  \hline
Total number\,$^a$, LF with $C\not= 1$& $1.1\times 10^{6}$& dashed line in Fig.~\ref{fig1}b\\ 			
QXMP, $C=1$ &  149 & solid line in Fig.~\ref{fig1}b \\
QXMP, $C\not= 1$ & {\bf 42} &  dashed line in Fig.~\ref{fig1}b\\
&12 --- 73 & uncertainties in Sect.~\ref{error}\\
\% QXMP in the survey&0.004&\\
\% QXMP in a volume$\,^b$&87&\\
\hline
Total number\,$^a$, varying $f_b$, $C\not= 1$ & $1.1\times 10^{6}$& red dashed line in Fig.~\ref{fig2}b \\			
QXMP, $C=1$ & 20 &  red solid line in Fig.~\ref{fig2}b \\
QXMP, $C\not= 1$ & {\bf 6} &  red dashed line in  Fig.~\ref{fig2}b \\
& 3 --- 12& uncertainties like in Sect.~\ref{error}\\
\% QXMP in the survey&0.0005&\\
\% QXMP in a volume$\,^b$&45&\\
\hline
Total number\,$^a$, exponential $f_b$, $C\not= 1$ & $1.1\times 10^{6}$& green dashed line in Fig.~\ref{fig2}b \\			
QXMP, $C=1$ & 24 &  green solid line in Fig.~\ref{fig2}b \\
QXMP, $C\not= 1$ & {\bf 7} &  green dashed line in  Fig.~\ref{fig2}b \\
& 4 --- 15& uncertainties like in Sect.~\ref{error}\\
\% QXMP in the survey&0.0006&\\
\% QXMP in a volume$\,^b$&42&\\
\hline
Total number\,$^a$, LF for centrals$\,^c$& $0.8\times 10^{6}$& solid line in Fig.~\ref{fig3}b \\			
QXMP, $C=1$ & 15 &  solid line in Fig.~\ref{fig3}b \\
QXMP, $C\not= 1$ & {\bf 4} &  dashed line in  Fig.~\ref{fig3}b \\
& 3 --- 9& uncertainties like in Sect.~\ref{error}\\
\% QXMP in the survey&0.0005&\\
\% QXMP in a volume$\,^b$&52&\\
\hline
\end{tabular}

\begin{tabular}{l}
\hskip 2.5cm $^a$\,The  real SDSS-DR7 spectroscopic survey has $0.93\times 10^{6}$ galaxies
\citep{2009ApJS..182..543A}.\\
\hskip 2.5cm $^b$\,In an unbiased, purely volume-limited survey.\\
\hskip 2.5cm $^c$\,From \citet{2009ApJ...695..900Y}.\\
\end{tabular}
\label{table1}
\end{centering}
\end{table*}

\subsection{Error budget for the expected number of QXMPs}\label{error}
The number of QXMPs depends on several assumptions, as explained in the 
previous section. Those assumptions are modified here to determine their impact on the estimated
number of QXMPs.

\begin{enumerate}
\item{\em Neglecting the scatter in the $SB_r$ versus $M_r$ relationship.}
The scatter increases the effective completeness quite substantially. 
If the scatter is neglected then one is left with the 
completeness function, $C$; either the actual completeness determined by 
\citet[][the symbols in Fig.~\ref{completeness}]{2005ApJ...631..208B} 
or the erf function  fit to this completeness (the smooth solid black line 
in Fig.~\ref{completeness}). If the actual completeness is used, 
then  $N_{\rm QXMP}\simeq 12$. If the erf fit is used, then  $N_{\rm QXMP}\simeq 19$.
These  estimates are closer to (but still larger than) the number of observed 
QXMPs (Eq.~[\ref{obs_qxmp}]).  However, they must be regarded as lower limits to 
the predicted $N_{\rm QXMP}$.   The scatter in the $SB_r$ versus $M_r$ relationship is a key 
ingredient of the detection process, as proven by the fact that the observed QXMPs
are high-SB outliers of the extrapolated $SB_r$ versus $M_r$ relation (Fig.~\ref{completeness}). 
Consequently, the scatter must be included, and  $N_{\rm QXMP} \geq 12$.
\item {\em The completeness function}. \label{item2}
We have used the erf fit to the completeness function by  \citet[][]{2005ApJ...631..208B}  
to evaluate the effective completeness leading to the limit in Eq.~(\ref{nqxmp}).   
If rather than the fit, the actual completeness is used (i.e., the symbols in Fig.~\ref{completeness}), 
then  $N_{\rm QXMP}\simeq 38$. The difference between the erf and the 
completeness by  \citet[][]{2005ApJ...631..208B}  is about $\pm 0.5$~mag, 
in the sense that if the erf fit is shifted by $\pm 0.5$~mag then it encompasses all the points.
If the effective completeness is evaluated using the completeness shifted 
by $\pm 0.5$~mag, one obtains the black dashed and dotted lines in Fig.~\ref{completeness}.
They yield a $N_{\rm QXMP}$ between 36 and 50.
\item  {\em The absolute magnitude limit.}\label{item3} The absolute magnitude limit to be an QXMP, 
given in Eq.~(\ref{faint3}), depends on a number of factors.
The value of this limit is relevant because the majority of the QXMPs are within
one magnitude of the cut-off (i.e., the number of QXMPs is dominated by objects
with metallicity just below the metallicity cut-off and its corresponding absolute 
magnitude). If we use exactly 1/10 
of the solar abundance by \citet{2009ARA&A..47..481A}, then the limit in 
 Eq.~(\ref{faint4}) becomes 7.69, so that the mass-metallicity 
relationship by \citet{2012ApJ...754...98B} predicts $M_B \geq -12.9$, and
the limit in the $r$-band becomes $M_r \geq -13.7$. This brighter limit allows
for more QXMPs, specifically, $N_{\rm QXMP}\simeq 73$.
The conversion from Eq.~(\ref{faint2}) to Eq.~(\ref{faint3}) employs both
the magnitude transformation by \citet{2005AJ....130..873J} and a single color 
$g-r\simeq 0.4 $ for all 
galaxies. If one uses the full range of colors for galaxies in the 
blue cloud, $0.2 \leq g-r \leq 0.6$ \citep[e.g.,][]{2009ARA&A..47..159B},
then $M_r$ goes from -13.0 to -13.5, which renders $N_{\rm QXMP}$ from
28 to 55.
Finally, if the uncertainties in the relation derived by
\citet{2012ApJ...754...98B} are propagated into the magnitude cutoff,
then $M_B \geq -12.5\pm 0.3$, and so  $M_r \geq -13.3\pm 0.3$, leading 
to values of $N_{\rm QXMP}$ between 28 to 64.
\item {\em The relationship between absolute magnitude and surface brightness.}\label{item4}
The relationship between the absolute magnitude and surface brightness
in Eq.~(\ref{mag_mag}) is given by \citet{2005ApJ...631..208B}.
In order to test the uncertainty introduced by the use of this relationship,
we also employed the law for blue objects by \citet[][]{2012AJ....143..102G}, $SB_r=29.9+0.46\,M_r$,
and by \citet{1985ApJ...295...73K}, $SB_r=30.3+0.47\,M_r$.
The latter was digitized from Fig.~3 in \citeauthor{1985ApJ...295...73K}'s paper. The 
magnitudes were then transformed from $V$ and $B$ to $r$ \citep{2005AJ....130..873J},
and the central surface brightness was converted to the half-light surface 
brightness assuming an exponential light profile. The use of these two alternative 
relationships renders a $N_{\rm QXMP}$ always around 42. 
\item {\em A combination of the previous assumptions.}
In the previous items, the ingredients that determine  $N_{\rm QXMP}$ are analyzed 
independently. We have also checked the combined effect of all of them operating
simultaneously. We carried out a Monte Carlo simulation where $N_{\rm QXMP}$ was
estimated varying the  completeness function, the magnitude limit, and the 
mapping between $SB_r$ and $M_r$, simultaneously. Explicitly, 
the center of the completeness function and the absolute magnitude limit
were randomly changed following Gaussian distributions with standard deviations of 
0.5\,mag (item~\#\,\ref{item2}) and 0.3\,mag (item~\#\,\ref{item3}), respectively.  
In addition, the  three relations between $SB_r$ and $M_r$ discussed
in item~\#\,\ref{item4} were assumed to be equally probable. As a result of 1000 trials, 
we  obtain $N_{\rm QXMP}= 46\pm 20$, which is similar to the range of values 
inferred from the individual factors separately (Table~\ref{table1}). 
\item {\em The adopted LF}. We separate the properties of the LF into two parts: the shape and 
the normalization. Changes in the shape are analyzed in Sect.~\ref{quenching}, and 
 produce significant changes in the number of expected QXMPs. The normalization, 
however, is fairly well-constrained by the total number of galaxies in the SDSS-DR7 spectroscopic
sample, which amount to  $0.93\times 10^6$ galaxies.  By
scaling the LF to reproduce the actual number of sources in the DR7 (i.e., to go from 1.1 million
to 0.93 million; see Table~\ref{table1}) one finds $N_{\rm QXMP}\simeq 35$.

\end{enumerate}



\subsection{Including galaxy formation quenching induced by the UV background}\label{quenching}
As we have shown above, the number of observed QXMPs (around 9; Eq.~[\ref{obs_qxmp}]) is 
not consistent with the number of QXMPs expected from extrapolating the observed LF to low 
luminosities (from 12 to 73, with the best value around 42;
Table~\ref{table1}).  Such an extrapolation of the LF implicitly neglects the 
quenching of galaxy formation in low-mass halos expected from numerical models of 
galaxy formation (see Sect.~\ref{motivation}).   These predict a rapid fall-off of the baryon 
fraction, $f_b$,  in low-mass halos due to various physical processes, such as 
heating of the intergalactic medium by the UV background or stellar feedback
(see Sect.~\ref{motivation}).
The drop in $f_b$ induces a drop in the gas that fuels star formation, 
and, hence, a drop in the number of the low-mass, low-luminosity galaxies
affected by the decrease of baryons. 

The effect of the baryon fraction on the LF can be modeled considering that
the magnitude of a  galaxy is related to the baryon fraction as follows,
\begin{equation}
M-M^\odot=-2.5\log\big[f_L\,(1-f_g)\,f_b\,\mu\big],
\label{defmag}
\end{equation}
where $M^\odot$, $f_L$, and $f_g$
stand for the absolute solar magnitude, 
the light-to-stellar mass ratio (in solar units), and the gas fraction, respectively. 
The symbol $\mu$ in Eq.~(\ref{defmag}) stands 
for the total mass, 
including dark matter, gas and stars. 
Equation~(\ref{defmag}) allows us to express $\Phi$ in terms of the LF obtained 
assuming the baryon fraction to be  constant, $\Phi_0$.
In this case, the mapping between $M$ and the magnitude $M_0$
when $f_b$ is a constant equal to $f_{b0}$ turns out to be
\begin{equation}
M-M_0=-2.5\log\big[f_b/f_{b0}\big],
\label{maineq1}
\end{equation}
so that the LFs for $\Phi(M)$ and $\Phi_0(M_0)$ are linked by \citep[e.g.,][]{martin71},
\begin{equation}
\Phi (M)=\Phi_0(M_0)\,{{dM_0}\over{dM}}.
\label{maineq0}
\end{equation}
The ratio between the two luminosity functions at the same magnitude, $X(M)$,
quantifies the drop in LF induced by the drop in the baryon 
fraction, i.e.,
\begin{equation}
X(M)={{\Phi(M)}\over{\Phi_0(M)}}={{\Phi_0(M_0)}\over{\Phi_0(M)}}\,{{dM_0}\over{dM}}.
\end{equation}
Neglecting variations with halo mass of the mass-to-light ratio and the gas
fraction, then
\begin{equation}
{{dM_0}\over{dM}}={{dM_0/d\mu}\over{dM/d\mu}}={1\over{1+d\ln f_b/d\ln\mu}}.
\label{maineq2}
\end{equation}

\begin{figure}
\includegraphics[width=0.49\textwidth]{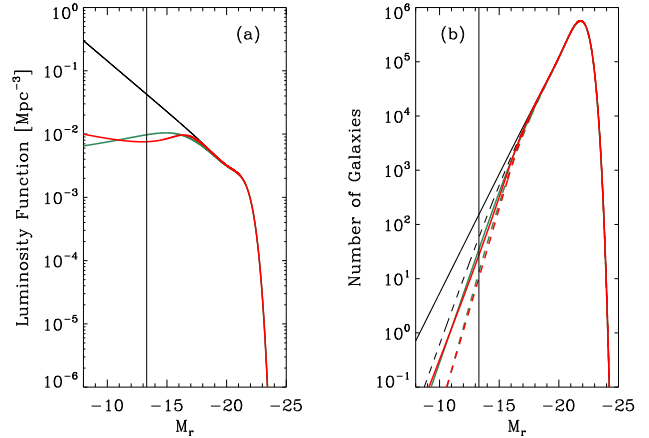}
\caption{(a) Galaxy LF in the $r$-band from 
\citet{2005ApJ...631..208B} (black solid line, which is identical 
to the black solid line in Fig.~\ref{fig1}a).
If the effect of the varying baryon fraction is considered, 
the LF becomes flat at low luminosities, as shown by the colored
lines. The red line corresponds to the variation computed 
by \citet{2008MNRAS.390..920O},  
whereas the green line represents an exponential drop.
(b)
Number of galaxies per unit magnitude to be expected from the 
SDSS-DR7 spectroscopic  survey. The color code is the same as in (a). 
The solid lines show the effect of the Malmquist bias 
(an apparent magnitude threshold), 
whereas the black dashed lines also include the incompleteness for 
low-SB objects.}
\label{fig2}
\end{figure}

The baryon fraction is usually expressed  in terms of the half-fraction mass, so that at 
mass $\mu= \mu_c$ the baryon fraction, $f_b$, is half the cosmic  baryon fraction
$\langle f_b\rangle$. In the parametrization determined by \citet{2000ApJ...542..535G},
and then  adopted  by many others, 
\begin{equation}
f_b=\langle f_b\rangle \big[1+(2^{a/3}-1)(\mu/\mu_c)^{-a}\big]^{-3/a},
\label{baryonf0}
\end{equation}
with $a\simeq 2$, as constrained by numerical simulations
\citep[][]{2008MNRAS.390..920O}.    
Numerical simulations also give $\mu_c\simeq 9.3\times 10^9\, {\rm M}_\odot$ in the local
Universe at redshift zero \citep[][Fig.~3]{2008MNRAS.390..920O}.
If the baryon fraction is given by Eq.~(\ref{baryonf0}), then the transformation between
$\Phi_0$ and $\Phi$ can be computed analytically, since
\begin{equation}
{{d\ln f_b}\over{d\ln \mu}}={{3(2^{a/3}-1)}\over{1+(2^{a/3}-1)(\mu_c/\mu)^a}}({{\mu_c}\over{\mu}})^a.
\label{derivative1}
\end{equation}  
It is important to realize that, in the limit of very low luminosities, 
$\mu\ll\mu_c$, so that Eq.~(\ref{derivative1}) predicts
\begin{equation}
{{d\ln f_b}\over{d\ln \mu}}\simeq 3.
\end{equation}  
Therefore, there is a drop of the LF associated with the vanishing baryon fraction, 
but it is not very large, 
\begin{equation}
\Phi(M){\Big/}\Phi_0(M_0)\simeq 0.25\,.
\label{limit0}
\end{equation}
In fact, $\Phi(M)$ flattens for very low masses  because, for a given variation of $M$,
$M_0$ changes very little, so that $\Phi_0(M_0)$ in Eq.~(\ref{limit0})
is approximately constant, and  $\Phi({\rm M})$ becomes independent of $M$ also.
The LFs in Fig.~\ref{fig2}a show this behavior -- see the red solid line, which is 
computed from 
Eqs.~(\ref{maineq0}),
(\ref{maineq1}),
(\ref{maineq2}),
(\ref{baryonf0}), and
(\ref{derivative1})
with $\langle f_b\rangle=0.158$  \citep{2016A&A...594A..13P}, 
$f_{b0}=\langle f_b\rangle$,
$f_L=1$, and $f_g=0.9$.

\begin{figure}
\includegraphics[width=0.48\textwidth]{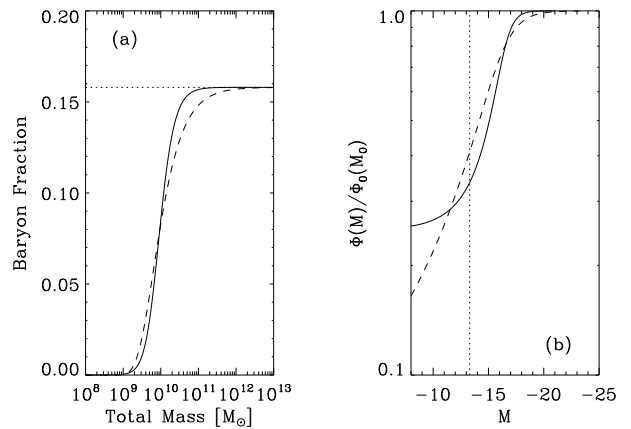}
\caption{Comparison between the two baryon  fractions used in this work. 
(a) Baryon fraction versus total mass
for the two parameterizations: the one by \citeauthor{2000ApJ...542..535G}
(the solid line) and the exponential drop (the black dashed line). In both 
cases, the half-fraction mass, $\mu_c$, has been set to 
$9.3\times 10^{9}\,{\rm M}_\odot$, with the universal baryon fraction 
given by $\langle f_b\rangle=0.158$ (the black dotted line). 
(b) Drop in the LF versus absolute magnitude produced by the two baryon fractions. 
The one by \citeauthor{2000ApJ...542..535G} saturates at 0.25 (the solid line),
whereas the exponential drop continues to decrease to infinity. 
Magnitudes are computed from total masses using the 
same parameters for the two baryon fractions, 
namely, a stellar mass-to-light ratio of 1 solar mass per solar luminosity, 
and a gas-to-stellar mass fraction of $f_g=0.9$. The vertical black dotted line denotes
the magnitude limit for a QXMP (Eq.~[\ref{faint3}]).
}
\label{fbvsfb}
\end{figure}

The damping of the LF caused by the baryon 
fraction in  Eq.~(\ref{baryonf0}) is never very large. In order to make the 
drop more pronounced, we also tried a negative exponential 
parametrization of the  baryon fraction,
\begin{equation}
f_b=\langle f_b\rangle\,{\Large\exp}(-{{\mu_c\ln 2}\over{\mu}}),
\label{baryonf1}
\end{equation}
which is hardly distinguishable from Eq.~(\ref{baryonf0}) in the 
representation used to compare with numerical simulations 
(see Fig~\ref{fbvsfb}a),
but which produces a linear drop of the luminosity 
function (Fig.~\ref{fbvsfb}b), since
\begin{equation}
{{d\ln f_b}\over{d\ln \mu}}=\mu_c\ln 2/\mu,
\end{equation}
so that, at low luminosity, where $\mu \ll \mu_c$,  
\begin{equation}
\Phi(M)/\Phi_0(M_0)\simeq {\mu/(\mu_c\ln 2)}\longrightarrow 0.
\label{limit1}
\end{equation} 
The LF resulting from the exponential fall-off of the baryon fraction is shown
as a green solid line in Fig.~\ref{fig2}a.

%
Figure~\ref{fig2}a is similar to Fig.~\ref{fig1}a, except that it 
reproduces the LFs when including the decrease  of baryon fraction 
towards low-mass halos. The red and the green lines correspond
to Eqs.~(\ref{baryonf0}) and  (\ref{baryonf1}), respectively, whereas the black solid line
is the same as in Fig.~\ref{fig1}a, and has been included for reference.
Figure~\ref{fig2}b shows the  number of galaxies expected in the 
SDSS-DR7 spectroscopic survey, considering only the apparent magnitude threshold 
(solid lines), and both the apparent magnitude threshold and the incompleteness (black dashed lines). 
In the case  of the baryon fraction given in Eq.~(\ref{baryonf0}), and considering the apparent 
magnitude limit and incompleteness, 
\begin{equation}
N_{\rm QXMP}\simeq 6,
\label{nqxmp1}
\end{equation}
which is significantly smaller than the estimate for the LF with the upturn at low
luminosity (Eq.~[\ref{nqxmp}]), and consistent with the observed number of QXMPs (Eq.~[\ref{obs_qxmp}]). 
The agreement with observations is enhanced even further after considering
the error budget expounded in the next paragraph.
The exponential baryon fraction (Eq.~[\ref{baryonf1}]) gives 
$N_{\rm QXMP}\simeq 7$, as  is reflected in Table~\ref{table1}.

Similar to the estimate in Eq.~(\ref{nqxmp}), the number in 
Eq.~(\ref{nqxmp1}) is quite uncertain. We have repeated the exercise 
in Sect.~\ref{error} for the case of LFs with varying baryon fraction.
The result is a  $N_{\rm QXMP}$ in the range between 3 and 12 
objects.
%
%
Another source of uncertainty, which we do not treat in Sect.~\ref{error} because
it does not affect Eq.~(\ref{nqxmp}), is the mapping between masses 
and magnitudes. According to Eqs.~(\ref{defmag}) and (\ref{baryonf0}), the
free parameters of this mapping,
$f_L$, $1-f_g$ and $\langle f_b \rangle$, appear in the equations as a single 
parameter, corresponding to their product. If this product is two times larger or 
smaller, $N_{\rm QXMP}$ changes from 5 to 8, respectively.
If, on the other hand, the half-baryonic fraction mass is 
varied from $2\times 10^{9}\,{\rm M}_\odot$ to
$2\times 10^{10}\,{\rm M}_\odot$, then    $N_{\rm QXMP}$ goes from 12 to 4.
(The nominal value we use is $9.3\times 10^{9}\,{\rm M}_\odot$.) 
These uncertainties in the estimate of  $N_{\rm QXMP}$ are summarized in Table~\ref{table1}.

%
\section{Discussion}\label{discussion}

\subsection{Are QXMPs central  or satellite galaxies?}\label{cen_or_sat}

The number of QXMPs in the SDSS-DR7 is not consistent with 
the extrapolation to low luminosities of the observed LF. Observations and 
theory agree much better if a varying baryon fraction is included to flatten  
the upturn of the observed LF at low luminosities, which implicitly assumes the 
QXMPs to be central galaxies of their dark matter halos. 
The baryon fraction of satellite galaxies is determined by interactions with 
nearby galaxies (tidal stripping and harassment), and with the circum-galactic 
medium of the central galaxy 
\citep[ram pressure stripping and starvation; e.g.,][]{2004IAUS..217..440C,2010PhR...495...33B}. 
Hence, the baryon fraction depends not only on the halo mass but 
on many other factors, and expressions for $f_b$ like Eq.~(\ref{baryonf0}) are 
no longer  valid. Consequently, the consistency of the estimated 
$N_{\rm QXMP}$ with observations suggests that the upturn in the 
observed LF is caused by the presence of satellites. 
This result is in agreement with the conclusion reached by 
\citet{2016MNRAS.459.3998L}. They model the LF by \citet{2005ApJ...631..208B} 
as the sum of a LF for centrals (i.e., the most massive galaxy in its 
dark matter halo) plus a LF for satellites (i.e., galaxies sharing a 
dark matter halo with other more massive galaxy). This decomposition reveals that the LF 
of field galaxies is dominated by satellite galaxies at $M_r > -17$, and that only halos 
more massive than $10^{10}$~M$_\odot$ contribute to the LF at $M_r < -12$.

\begin{figure}
\includegraphics[width=0.49\textwidth]{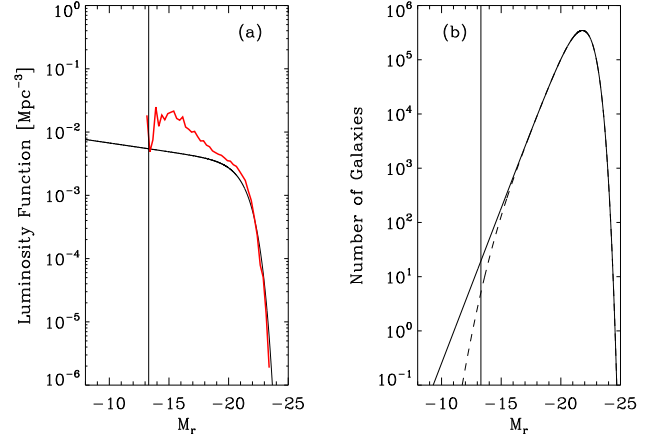}
\caption{(a) LF in the $r$-band for central galaxies, determined by \citet{2009ApJ...695..900Y}. The red line shows the LF by 
\citet{2005ApJ...631..208B}; it is the same as the red line in Fig.~\ref{fig1}a,
and has been included here for reference. 
(b) Number of galaxies per unit magnitude to be expected in the SDSS-DR7 
spectroscopic survey for the LF in (a). 
The solid line shows the effect of the Malmquist bias, whereas the black dashed 
line also includes the incompleteness for low-SB objects.
}
\label{fig3}
\end{figure}

The conjecture that QXMPs are central galaxies rather than satellite galaxies  
is also qualitatively consistent with the observed $N_{\rm QXMP}$.
If one uses the empirical LF for central galaxies determined  by  
\citet{2009ApJ...695..900Y},
and simply extrapolates it to low luminosity,
then $N_{\rm QXMP}\simeq$\,3 -- 9; see Figs.~\ref{fig3}a and \ref{fig3}b, and Table~\ref{table1}.
This LF for centrals does not show the upturn and stays below the LF by 
\citet{2005ApJ...631..208B} (see Fig.~\ref{fig3}). The lack of an
upturn produces  a major drop in the number of QXMPs. The LF for
centrals also predicts 30\,\% less galaxies than the reference LF 
by \citet{2005ApJ...631..208B}. This difference is due to low-luminosity
($M_r > -17$)  satellite galaxies, included by \citet{2005ApJ...631..208B},
but separated out  by  \citet{2016MNRAS.459.3998L}.

\subsection{Factors limiting the number of observed QXMPs}\label{exploring_limits}

Assuming that our model for the selection function
provides a good representation of the SDSS properties, 
one can investigate which, among the parameters of the survey,
are responsible for the limited number of observed QXMPs. 
In principle, there are three main parameters that control
such a number, namely, (1) the completeness, (2) the 
area coverage of the survey, and (3) the apparent magnitude 
limit.

The number of objects of a particular type in the 
survey linearly scales with the area of the survey. Since 
SDSS already covers a significant part of the sky
\citep[around 20\,\%;][]{2009ApJS..182..543A}, this is not a
limiting factor in the present case. In other words,  no dramatic (tenfold) increase 
in the number of QXMPs will follow from increasing the area covered 
by the SDSS-DR7.

\begin{figure}
\includegraphics[width=0.49\textwidth]{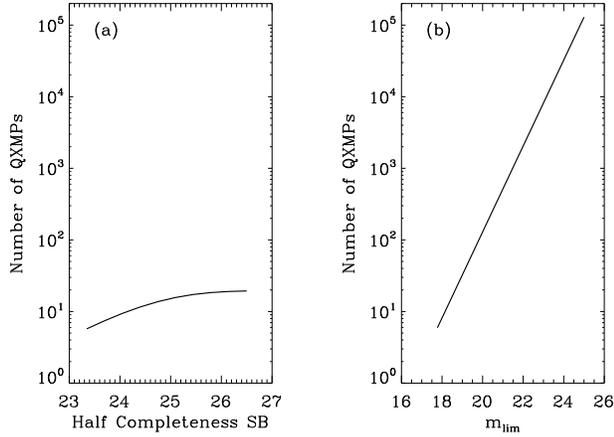}
\caption{
(a) Number of QXMPs expected when the SB cutoff of the completeness function 
changes.  The cutoff is quantified as the SB having a completeness  of one 
half.  It only has a moderate impact on $N_{\rm QXMP}$.
(b) Number of QXMPs expected in the SDSS-DR7 spectroscopic 
survey if the apparent magnitude limit, $m_{lim}$, could be modified. 
This parameter critically determines $N_{\rm QXMP}$.
The range of the ordinates in (a) and (b) is the same.
}
\label{future}
\end{figure}

Even though completeness is an important factor, it is not the key 
factor that determines $N_{\rm QXMP}$. 
At the  absolute magnitude limit that characterizes the QXMPs, 
the completeness is around 30\,\%\ (see Fig.~\ref{completeness}).
Although the completeness can be increased, it cannot exceed the 
value of one, which in turn implies a moderate increase in $N_{\rm QXMP}$.
This dependence can be apprised in Fig.~\ref{future}a, where  
the expected number  of QXMPs is represented for 
different SB cutoffs of the completeness function. To compute 
the number, we have used 
the same completeness of the SDSS-DR7 (Fig.~\ref{completeness}, the 
red solid line), shifted in $SB_r$, with the shift 
parameterized as the one-half completeness $SB_r$.  
As the SB of the drop  increases,  $N_{\rm QXMP}$ increases
too. However, it saturates at a value that is only ten times
larger than the value corresponding to the SDSS-DR7 spectroscopic 
survey (the point of lower SB in Fig.~\ref{future}a).
The curves in  Fig.~\ref{future}a and \ref{future}b assume 
the LF shown as a red solid line in Fig.~\ref{fig2}a.

The apparent magnitude limit turns out to be the critical parameter
that limits the number of QXMPs. It determines
the volume sampled by the survey, which, in principle, can be increased
indefinitely as $m_{limit}$ increases. The behavior is shown 
in  Fig.~\ref{future}b. Just to provide an idea of the increase,
if one consideres the apparent magnitude limit of the 
SDSS-DR7 {\em photometric} survey ($m_{limit}=22.2$ in the $r-$band), then
Fig.~\ref{future}b predicts the presence of approximately 2800 QXMPs,
which should be compared to the prediction of 6 QXMPs in the
SDSS-DR7 {\em spectroscopic} sample ($m_{limit}=17.77$;
see Table~\ref{table1} and Fig.~\ref{future}b). 

\subsection{Number of QXMPs in other surveys}\label{other_surveys}

The formalism developed in Sect.~\ref{estimate1} allows us to estimate the 
number of expected QXMPs in any magnitude-limited survey, provided their
magnitude limit, half-completeness, $SB_r$, and coverage area.
If the completeness function is assumed to have the functional form of the 
SDSS completeness, then the half-completeness, $SB_r$, fully describes it.
We carried out the exercise of estimating $N_{\rm QXMP}$ for a number of ongoing and 
forthcoming  large area surveys, specifically, for the 
the Dark Energy Survey  (DES; The DES Collaboration~\citeyear{2005astro.ph.10346T}),
the  Galaxy and Mass Assembly survey \citep[GAMA;][]{2015MNRAS.452.2087L},
the Kilo-Degree Survey  \citep[KIDS;][]{2015A&A...582A..62D},
and the Large Synoptic Survey Telescope \citep[LSST;][]{2008arXiv0805.2366I}.
%
%
\begin{table*}
\centering 
\caption{$N_{\rm QXMP}$ for various ongoing and forthcoming galaxy surveys}
\begin{tabular}{lccccc}
\hline
\noalign{\smallskip} 
Survey  & $r$-band~$m_{lim}$ &Half-Completeness SB& Area &$N_{\rm QXMP}\,^a$ & $N_{\rm QXMP}\,^b$\\
& [mag] &  [mag~arcsec$^{-2}$]&[deg$^2$]&&\\
\noalign{\smallskip} 
 \hline
 \noalign{\smallskip} 
SDSS$\,^c$ spectroscopic & 17.8 & 23.4&8032&42&6\\
GAMA$\,^d$&19.8&26.0&286&83&11\\
SDSS$\,^c$ photometric & 22.2 & 23.4&8423&2.0$\times 10^{4}$&2.8$\times 10^{3}$\\
KIDS$^e$&24.0&27.2&1500&1.5$\times 10^5$&2.0$\times 10^4$\\
DES$\,^f$&24.1&27.1&5000&5.8$\times 10^5$&7.6$\times 10^4$\\
LSST$\,^g$, single visit&24.7&25.9&18000&4.3$\times 10^6$&6.0$\times 10^5$\\
LSST$\,^g$, co-added 10 year&27.5&28.7&18000&2.3$\times 10^8$&3.0$\times 10^7$\\
\noalign{\smallskip} 
 \hline
\end{tabular}
\begin{tabular}{l}
\noalign{\smallskip} 
 $^a$~Assuming a LF function growing with increasing magnitude (the black solid line in Fig.~\ref{fig2}a).\\
$^b$~Assuming a LF function decreasing with increasing magnitude (the red solid line in Fig.~\ref{fig2}a).\\
$^c$~For DR7. Parameters from the webpage: {\tt http://classic.sdss.org/dr7/}.\\
$^d$~Galaxy and Mass Assembly. Parameters from  \citet{2015MNRAS.452.2087L}.\\
$^e$~Kilo-Degree Survey. Parameters from {\citet{2015A&A...582A..62D}}\\
$^f$~Dark Energy Survey, after 5 years. Parameters from  The DES Collaboration~(\citeyear{2005astro.ph.10346T}).\\
$^g$~Large Synoptic Survey Telescope. Parameters from  {\citet{2008arXiv0805.2366I}}.\\
\end{tabular}
\label{tab:surveys}
\end{table*}
The parameters that define the surveys are given in Table~\ref{tab:surveys}. 
LSST, DES and KIDS do not explicitly give the half-compleness, SB. In these cases,
we infer it from the depth for point sources as described in Appendix~\ref{appb}.

As is shown in Table~\ref{tab:surveys}, the number of QXMPs expected in the next generation
of wide area surveys is very large, reaching up to $10^7$  in the 10-year average LSST.
The actual number changes by one order of magnitude, depending on whether the LF 
increases towards low luminosity (the black solid line in Fig.~\ref{fig2}a) or if it flattens 
out  (the red solid line in Fig.~\ref{fig2}a). Therefore, it should be easy to discriminate both 
trends using these new surveys. For example, the ongoing GAMA survey predicts 10 or 100 QXMPs,
depending on the LF faint end.  In agreement with the conclusion in the previous section, 
the key factor  determining $N_{\rm QXMP}$ is not so much the SB cutoff but the limiting magnitude, $m_{lim}$. 
Note, however, that even if the surveys contain all these new QXMPs, it will
be imposible to confirm the XMP nature of many of the faint objects. The spectroscopic 
follow up required to determine abundances will be possible only in those QXMPs 
where star-forming regions are bright enough.

%
\section{Conclusions}\label{conclusions}

Galaxies follow a relationship between luminosity and gas-phase metallicity,  
so that faint  galaxies tend to be metal-poor galaxies as well. Since the 
luminosity  function increases steeply towards low luminosity, one would
 naively expect that most observed galaxies are metal-poor. This is not the
case. This apparent inconsistency is usually attributed to the 
low-luminosity of the metal-poor objects, which are under-represented 
in galaxy surveys. Firstly, this occurs due to the Malmquist bias: 
surveys are apparent magnitude-limited, so that the sampled 
volume drops down dramatically for faint sources.
Secondly, low-luminosity galaxies are also low-SB
galaxies, and the surveys tend to miss extended, low-SB objects. 

This dearth of metal-poor galaxies is particularly severe for the so-called
extremely metal poor (XMP) galaxies, with a  gas-phase metallicity smaller than 
a tenth of the solar metallicity. They are of astrophysical interest for 
a number of reasons highlighted in Sect.~\ref{motivation}, but they 
represent only a tiny fraction of the galaxies in the most 
popular surveys 
\citep[e.g., 0.02\,\% in the recent SDSS-DR7 search by ][]{2016ApJ...819..110S}. 
Moreover, most of the observed XMPs are outliers of the luminosity-metallicity relationship. Therefore, they are not part of the predicted
sea of faint  XMPs. (We denote the faint XMPs as {\em quiescent} XMPs or 
QXMPs.) 
The question arises as to whether the actual number of observed
XMPs is {\em quantitatively} consistent with the expected number. 
We address this question in the present paper. Most known XMPs 
come from the  SDSS-DR7 spectroscopic survey, so
we compare the number of observed QXMPs in this 
survey with the expected number. The main conclusion
of our work is that they disagree, unless the luminosity 
function for QXMPs is considerably shallower than the extrapolation
to low luminosity of the observed LF. 

The number of QXMPs in the SDSS-DR7 spectroscopic survey 
turns out to be $9\pm 3$, with the error bar representing the
Poissonian fluctuation (Sect.~\ref{obs_constraints}, Table~\ref{coordinates}, 
and Figs.~\ref{observation} and \ref{images}). 
Extrapolating to $M_r > -13.3$
the LF for faint galaxies observed by \citet[][shown in Fig.~{\ref{fig1}a}]{2005ApJ...631..208B}, 
the expected number is 42 (Sect.~\ref{estimate1}).  
This estimate takes into account the Malmquist bias plus 
the finite completeness of SDSS for low-SB objects. 
In addition, it includes the scatter in the SB versus magnitude 
relationship, that effectively increases the completeness of the survey
at low luminosities (Appendix).  Once the various uncertainties involved in the estimate are 
considered  (Sect.~\ref{error}), the expected number of QXMPs is in the range between 
12 and 73,  with the low value highly disfavored. 

On the other hand, if the previous LF is modified to include the decrease of  
baryon fraction in low-mass dark matter halos, then 
the upturn at low luminosity disappears (Fig.~\ref{fig2}a), rendering 
an expectation of only 6 QXMPs. (The number is between 3 and 12 when 
uncertainties  are taken into account; see Sect.~\ref{quenching}.) 
The tension with observation automatically disappears. Including the varying 
baryon fraction implicitly assumes the QXMPs to be central galaxies in their 
dark matter haloes. In fact, the LF for centrals determined by 
\citet[][see Fig.~\ref{fig3}a]{2009ApJ...695..900Y} 
does not show the upturn, and it predicts between 3 and 9 QXMPs
(Sect.~\ref{cen_or_sat} and Table~\ref{table1}). 
The agreement with observations has several implications.
Firstly, QXMPs seem to be centrals, rather than satellite galaxies. 
QXMPs become  tracers of low-mass halos not 
gravitationally bound to more massive halos, and so they can be 
used to trace these haloes observationally. These low-mass dark matter halos 
are of clear astrophysical interest in the context of characterizing
the building blocks in the hierarchical formation of galaxies and,
in particular, the effect of the cosmic UV background in their 
baryonic content.
The fact that QXMPs seem to be centrals is consistent with
the observation that most XMPs appear to be isolated and in low density
regions of the Universe \citep[e.g.,][]{2015ApJ...802...82F,2016ApJ...819..110S}.
Secondly, the upturn in the faint end of the observed LF appears to be 
due to satellite galaxies, and this should be taken into account 
when comparing observations and numerical models. 
Thirdly, the baryon fraction predicted by the numerical
models by \citet{2008MNRAS.390..920O} is consistent with observations.
Finally, there is no expectation of finding a significant number
of new QXMPs in the SDSS-DR7 spectroscopic survey. 

Assuming that our modeling of the SDSS biases 
is correct, we have studied which, among  the parameters defining the survey, 
restricts the number of QXMPs most  (Sect.~\ref{exploring_limits}). It turns out to 
be the apparent magnitude limit, which  is more relevant than the 
incompleteness. Thus, the photometric SDSS-DR7 survey, which is 4 magnitudes 
deeper than the spectroscopic one, should contain as many as 2800 QXMPs.  
The expected numbers in various other surveys are determined in Sect.~\ref{other_surveys},
and a summary is presented in Table~\ref{tab:surveys}.
Future surveys are predicted to detect QXMPs in large quantities.  

\acknowledgements
%
%
Thanks are due to an anonymous referee for helping us to reinforce
the error budget worked out in Sect.~\ref{error}. This
work has been partly funded by the Spanish Ministry of Economy and 
Competitiveness (MINECO), projects {\em Estallidos\,5} AYA2013--47742--C04--02--P
and {\em Estallidos\,6} AYA2016-79724-C4-2-P.
MEF gratefully acknowledges the financial support of {\em Funda\c c\~ao para
a Ci\^encia e Tecnologia} (FCT -- Portugal), through the research grant 
SFRH/BPD/107801/2015, and of the  {\em Estallidos} project.
CDV acknowledges financial support from MINECO through grants AYA2013-46886 and
AYA2014-58308, and under the Severo Ochoa Programs SEV-2011-0187 and SEV-2015-0548.
EDS thanks a Severo Ochoa fellowship at the IAC which led to participation in 
this research.
This research has made use of NASA's Astrophysics Data System
Bibliographic Services and the NASA/IPAC Extragalactic Database
(NED), which is operated by the Jet Propulsion Laboratory, California
Institute of Technology, under contract with NASA.
%
%
\newcommand\jcap{JCAP}
\newcommand\pasa{PASA}

\appendix
\section{Scatter in the $SB$ versus $M$ relationship}\label{appa}

Equation~(\ref{select1}) assumes a one-to-one correspondence 
between the SB and the absolute magnitude 
of the source. The fact that most observed QXMPs are high-SB
outliers of the $SB$ versus $M$ relationship indicates that
the scatter in this relationship may be 
of importance in our estimate.  In simple terms, an object 
is preferentially picked out by SDSS if it is of high SB for 
its absolute magnitude.

In order to treat this case in our formalism, we assume the 
LF to be the marginal probability density  \citep[see, e.g.,][]{martin71}
of a  bi-variate joint probability density function (PDF), $P(M,SB)$, that 
quantifies the number of galaxies with
absolute magnitude $M$ and surface brightness $SB$, i.e.,
\begin{equation}
\Phi(M)=\int_{\forall\,SB} P(M,SB)\,dSB.
\end{equation}  
Then computing the number is equivalent to Eq.~(\ref{eq1}), but it involves
a double integral over $M$ and $SB$,
\begin{equation}
N(M>M_{lim})=\int^{M_1}_{M_{lim}}\int_{\forall SB}\,P(M,SB)\,V(M)\,C'(SB)\,dSB\,dM,
\label{newtotal}
\end{equation} 
with the symbol $C'$ standing for the completeness function
expressed in terms of the surface brightness. The bi-variate 
distribution function can be expressed in terms of the conditional 
probability function, $P(SB|M)$ (the probability of having a 
surface brightness $SB$, given that the absolute magnitude is $M$), 
\begin{equation}
P(M,SB)=P(SB|M)\,\Phi(M).
\label{bivariate}
\end{equation}
Inserting the expression~(\ref{bivariate}) into Eq.~(\ref{newtotal})
one recovers Eqs.~(\ref{eq1}) and (\ref{select1}), provided that 
$C(M)$ is replaced with an effective  completeness function, $C_{eff}$, given by
\begin{equation}
C_{eff}(M)=\int_{\forall\,SB} P(SB|M)\,C'(SB)\,dSB.
\label{ceff}
\end{equation}
With minimal assumptions, one can write down the conditional 
probability density function as  
\begin{equation}
 P(SB|M)={{1}\over{\Delta_0}}\,G({{SB-SB_0}\over{\Delta_0}}),
\end{equation}
with 
\begin{displaymath}
SB_0=SB_0(M),
\end{displaymath}
\begin{displaymath}
\Delta_0=\Delta_0(M),
\end{displaymath}
and $G$ any positive function properly normalized,
\begin{equation}
\int_{-\infty}^{+\infty}G(x)\,dx=1.
\end{equation} 

The case without scatter in the SB versus M relationship corresponds to an
infinitely narrow conditional probability function, which
we can easily treat in the limit $\Delta_0\longrightarrow 0$,
so that 
\begin{equation}
{{1}\over{\Delta_0}}\,G\big({{SB-SB_0}\over{\Delta_0}}\big) \longrightarrow 
\delta(SB-SB_0),
\end{equation} 
with $\delta$ a delta Dirac function. 
Inserting the previous expression into Eq.~(\ref{ceff}) yields
\begin{equation}
C_{eff}(M)=C(M)=C'(SB_0(M)),
\end{equation}
with $SB_0(M)$ given by expression~(\ref{mag_mag}). In general,
the computation of $N_{\rm QXMP}$  must make use of the full expression, where the 
{\it effective} completeness function is a type of convolution of the completeness 
function with the conditional PDF. Following \citet{2005ApJ...631..208B}, we will assume 
that the conditional PDF is a Gaussian,
\begin{equation}
G(x)={{1}\over{\sqrt{2\pi}}}{\Large \exp}(-{{x^2}\over{2}}),
\end{equation}  
with the mean, $SB_0$, and
the variance, $\Delta^2_0(M)$, measured to be 
\begin{equation}
SB_0(M)=23.8+0.45\,(M+13.3), 
\label{mag_mag2}
\end{equation}
\begin{displaymath}
\Delta_0(M)=1.16+0.081\,(M+13.3).
\end{displaymath}
Equation~(\ref{mag_mag2}) is identical to Eq.~(\ref{mag_mag}), and has been 
included here for comprehensiveness. The effective completeness function 
resulting from $C'$ in \citet{2005ApJ...631..208B} and the 
above parametrization of the conditional function  is shown as the red solid line 
in Fig.~\ref{completeness}.  The net effect is a significant enhancement 
of the completeness at low luminosity.

We note that in the limiting case when (1) $C'(SB)$ is a Heaviside step function (zero for
$SB$ larger than a given surface brightness, and one elsewhere), and when (2) $G$ 
is a Gaussian of constant width $\Delta_0$, then $C_{eff}$ in Eq.~(\ref{ceff}) 
has the shape of an erf function. This approximation is used in the main text.


\section{Estimate of the SB limit from the depth of the survey}\label{appb}
In order to predict the number of QXMPs expected in various surveys,
one needs to assign a SB limit to them.
The requirements of the survey are often set in terms of the {\rm depth} for point sources,
i.e., the flux that a point source must have to grant detection with a signal-to-noise ratio
a number of times above the noise level. In order to transform this 
parameter to the corresponding SB limit,  
we proceed as follows:  the {\em depth} for point sources, $m_{\rm point}$, is defined
as,
\begin{equation}
m_{\rm point}=-2.5\,\log(\xi\,\sigma\sqrt{n_{\rm point}})+\kappa,
\label{app2:eq1}
\end{equation} 
where $\sigma$ stands for the noise per pixel, $\xi$ for the level above noise to grant 
detection, and $n_{\rm point}$ for the number of pixels covered by a point source.
$\kappa$ sets the zero of the magnitude scale. Equation~(\ref{app2:eq1}) assumes
the noise of adjacent pixels to be independent, so that the noise of the sum of them 
adds up quadratically. We will define the SB limit as 
the SB of an extended source exceeding the noise in one arcsec, i.e.,
\begin{equation}
m_{\rm ext}=-2.5\,\log(\sigma\sqrt{n_{\rm 1''}})+\kappa,
\label{app2:eq2}
\end{equation} 
 with  $n_{\rm 1''}$ for the number of pixels in one arcsec. Since
\begin{equation}
n_{\rm point}/n_{\rm 1''}=\pi\,{\rm (FWHM/2)^2},
\end{equation}
with FWHM the full width half-maximum of the point-spread funcion
in arcsec, then Eqs.~(\ref{app2:eq1}) and (\ref{app2:eq2})
lead to,
\begin{equation}
m_{\rm ext}=m_{\rm point}+2.5\log(\xi{{\sqrt{\pi}}\over{2}}{\rm FWHM}).
\label{app2:eq3}
\end{equation}
Equation~(\ref{app2:eq3}) links the depth, $m_{\rm point}$, with the SB limit, $m_{\rm ext}$. 

\end{document}